\documentclass[letterpaper,12pt]{article}
\usepackage[dvips]{graphicx}
\usepackage{amsmath}
\usepackage{subfigure}
\usepackage{psfrag} 

\newlength{\dinwidth}
\newlength{\dinmargin}
\numberwithin{equation}{section}
\def\be{\begin{equation}}
\def\ee{\end{equation}}
\def\bea{\begin{eqnarray}}
\def\eea{\end{eqnarray}}
\def\la{\langle}
\def\ra{\rangle}

\newbox\charbox
\newbox\slabox
\def\s#1{{
        \setbox\charbox=\hbox{$#1$}
        \setbox\slabox=\hbox{$/$}
        \dimen\charbox=\ht\slabox
        \advance\dimen\charbox by -\dp\slabox
        \advance\dimen\charbox by -\ht\charbox
        \advance\dimen\charbox by \dp\charbox
        \divide\dimen\charbox by 2
        \raise-\dimen\charbox\hbox to \wd\charbox{\hss/\hss}
        \llap{$#1$}
}}

\begin{document}
\titlepage
\begin{flushright}
IPPP/06/15 \\
DCPT/06/30 \\
March 2006 \\

\end{flushright}

\vspace*{0.5cm}

\begin{center}
{\Large \bf Scattering amplitudes with massive fermions using BCFW recursion}

\vspace*{1cm}
\textsc{K.J. Ozeren and W.J. Stirling} \\

\vspace*{0.5cm} Institute for
Particle Physics Phenomenology, \\
University of Durham, DH1 3LE, UK \\
\end{center}

\vspace*{0.5cm}

\begin{abstract}
\noindent We study the QCD scattering amplitudes for $\bar{q}q \to gg$ and $\bar{q}q \to ggg$ where $q$ is a massive fermion. Using a particular choice of massive fermion spinor we are able to derive very compact expressions for the partial spin amplitudes for the $2\to 2$ process. We then investigate the corresponding $2\to 3$ amplitudes using the BCFW recursion technique.
For the helicity conserving partial amplitudes we again derive very compact expressions, but were unable to treat the helicity-flip amplitudes recursively, except for the case where all the gluon helicities are the same. We therefore evaluate the remaining partial amplitudes using standard Feynman diagram techniques.
\end{abstract}

\newpage

\section{Introduction}
In the last two years there has been dramatic progress in the calculation of multi-particle scattering amplitudes in quantum field theory. Following Witten's \cite{Witten} discovery of a connection between QCD amplitudes and twistor string theory, a calculational technique \cite{CSW} was found which has come to be known as `the CSW construction'. It amounts to an effective scalar perturbation theory, in which MHV amplitudes are elevated to the status of vertices, connected by scalar propagators. This scheme found wide application \cite{MHV1,MHV2,MHV3,MHV4,MHV5,MHV6}, though it turned out that there was an even more efficient way to calculate scattering amplitudes. Britto, Cachazo, Feng and Witten found a recursion relation \cite{BCF,BCFW} which, by shifting momenta, takes advantage of the analytic properites of tree amplitudes. Use of the BCFW recursion relation led easily to very compact expressions. Originally applied to purely gluonic tree amplitudes, the recursion has since been extended to include fermions \cite{LuoWen,KJO}, gravitons \cite{gravity} and loop amplitudes \cite{loops1,loops2,loops3,loops4}. As well as perhaps giving hints of as yet unknown mathematical structure beyond the Standard Model, these developments are potentially important for the calculation of Standard Model backgrounds at colliders such as the LHC. The more accurately the relevant cross sections are known, the higher the discovery potential of the machine will be.

One area of very recent progress is the calculation of amplitudes involving massive fermions. It was shown in Ref.~\cite{Schwinn} how to generalize Supersymmetric Ward Identities \cite{SWI} to include massive particles. In this way, different amplitudes involving fields belonging to the same supersymmetric multiplet are related by a rotation. For instance \cite{Ferrario}, amplitudes involving quarks and gluons are related by SWIs to amplitudes involving scalars and gluons, and these have been calculated in Ref.~\cite{scalars}. The off-shell Berends-Giele \cite{Berends} recursion has also proved useful \cite{Rodrigo}. Tree amplitudes with massive fermions are required as input within the unitarity \cite{unitarity} method to calculate one-loop amplitudes, and to this end Ref.~\cite{rozali} provides four- and five-point amplitudes with D-dimensional fermions, calculated using BCFW recursion.

The recursion relations were extended in Ref.~\cite{Badger1} to include massive fermions, and in \cite{Badger2} four-point amplitudes involving two massive quarks and two gluons were calculated. The topic of the present work is to extend this to
five-point amplitudes, using a treatment of massive fermion spinors introduced some twenty years ago in Ref.~\cite{KS2}. We first outline the particular spinor helicity method we will use, and then we show that $2 \to 2$ scattering processes in QCD can be written in a form which is ideally suited for use in BCFW recursion. In Section 4 we use the recursion relations to derive compact expressions for certain $\bar{q} q \to ggg$ partial amplitudes. Those partial amplitudes which we could not treat recursively are evaluated using Feynman diagrams in Section 5. Finally, we present our conclusions.

\section{Spinor Products}

For many years amplitudes involving massless momenta $p_i$ and $p_j$ have been expressed in terms of spinor products,

\be
[ij] = \bar{u}^+(p_i) u^-(p_j) \hspace{5mm} \mbox{and} \hspace{5mm} \la ij \ra = \bar{u}^-(p_i) u^+(p_j).
\ee
In this way amplitudes find their simplest expression. The spinors in question can be thought of either as two-component Weyl or 4-component Dirac spinors. Numerical evaluation of such amplitudes involves the use of the standard formulae for the spinor products in terms of the momentum 4-vectors. Following \cite{KS2}, let us first introduce two 4-vectors $k_0$ and $k_1$ such that
\be
k_0\cdot k_0 = 0 \hspace{2mm},\hspace{2mm} k_1\cdot k_1 = -1 \hspace{2mm},\hspace{2mm} k_0\cdot k_1 = 0.
\ee
We now define a basic spinor $u_-(k_0)$ via
\be
u^-(k_0)\bar{u}^-(k_0) = \frac{1 - \gamma^5}{2} \s{k}_0,
\ee
and choose the corresponding positive helicity state to be
\be
u^+(k_0) = \s{k}_1 u^-(k_0).
\ee
Using these definitions it is possible to construct spinors for any null momentum $p$ as follows:
\be \label{defineP}
u^{\lambda}(p) =  \frac{\s{p} \ u^{-\lambda}(k_0)}{\sqrt{2p\cdot k_0}},
\ee
with $\lambda = \pm$. Note that this satisfies the massless Dirac equation $\s{p}u(p) = 0$, as required. We can now simply evaluate spinor products. For example,
\bea
[ij] &=& \bar{u}^+(p_i) u^-(p_j) \\
&=& \frac{\bar{u}^-(k_0) \ \s{p_i} \ \s{p_j} \ u^+(k_0)}{\sqrt{4 \ (p_i\cdot k_0)(p_j \cdot k_0)}} \\
&=& \frac{\mbox{Tr}(\frac{(1-\gamma^5)}{2} \ \s{k_0} \ \s{p_i} \ \s{p_j} \ \s{k_1})}{\sqrt{4 \ (p_i\cdot k_0)(p_j \cdot k_0)}} \\
&=& \frac{(p_i \cdot k_0) (p_j \cdot k_1) - (p_j \cdot k_0)(p_i \cdot k_1) - i \epsilon_{\mu \nu \rho \sigma}k_0^\mu p_i^\nu p_j^\rho k_1^\sigma}{\sqrt{(p_i\cdot k_0)(p_j \cdot k_0)}}.
\eea
A similar expression is obtained for the angle product $\la ij \ra$. The arbitrary $k_0$ and $k_1$ can now be chosen so as to yield as simple an expression for the product $[ij]$ and $ \la ij \ra$ as possible, to facilitate numerical evaluation of the ampitudes. The choice\footnote{The notation is $k^\mu = (k^0,\mbox{\bf k})$.}
\bea \label{define k}
k_0 = (1,0,0,1) \\
k_1 = (0,0,1,0)
\eea
is a good one, giving the familiar
\be
[ij] = (p_i^y + ip_i^x) \left[ \frac{p_j^0 - p_j^z}{p_i^0 - p_i^z} \right]^{\frac{1}{2}} - (p_j^y + ip_j^x) \left[ \frac{p_i^0 - p_i^z}{p_j^0 - p_j^z} \right]^{\frac{1}{2}}.
\ee
\subsection{Massive spinors}
To evaluate spinor products involving massive spinors, we need to find a definition analogous to \eqref{defineP}. One possibility is that outlined in \cite{KS2},
\be \label{defineMassiveP}
u^{\lambda}(p) = \frac{(\s{p} + m) u^{-\lambda}(k_0)}{\sqrt{2 p\cdot k_0}},
\ee
which satisfies the massive Dirac equation, $(\s{p} - m)u^{\lambda}(p) = 0$. The $m$ in \eqref{defineMassiveP} is positive or negative when $u^{\lambda}(p)$ describes a particle or antiparticle respectively. This definition has the virtue\footnote{Care is needed when $p \cdot k_0$ also vanishes in this limit, as we will discuss later.} of being smooth in the limit $m \to 0$. We will use \eqref{defineMassiveP} to evaluate products involving massive spinors.

It is easily seen that the familiar $[. \ .]$ and $\la . \ . \ra$ products take the same form for massive spinors as they do for massless ones. Explicit mass terms drop out due to various trace theorems. However, the product of like-helicity spinors is now non-zero:
\bea
(ij) &=& \bar{u}^{\pm}(p_i) u^{\pm}(p_j), \\
&=& m_i \left(\frac{p_j \cdot k_0}{p_i \cdot k_0}\right)^{\frac{1}{2}} + i \leftrightarrow j\\
\label{round bracket} &=& m_i \left(\frac{ p_j^0 - p_j^z }{p_i^0 - p_i^z}\right)^{\frac{1}{2}} + i \leftrightarrow j,
\eea
where in the last line we have used $k_0$ as given in \eqref{define k}. Note that the like-helicity product is the same whatever the helicity of the spinors involved, and that we use a round bracket as a shorthand notation for it.

We have been using the word `helicity' to refer to the spin projection of massive fermions, but in fact this is only justified if the projection is onto the direction of the momentum vector, and it is not obvious that this is the case. There exists a unique polarization vector, though it depends on the arbitrary $k_0$,
\be
\sigma ^{\mu} = \frac{1}{m} \left(p^{\mu} - \frac{m^2}{p \cdot k_0} k_0^{\mu} \right).
\ee
The spinors \eqref{defineMassiveP} satisfy
\be
\big(1-\lambda \gamma^5 \s{\sigma}\big) u^{\lambda} = 0.
\ee
We see that besides the momentum $p$ there is an additional contribution to the polarization vector proportional to $k_0$. Suppose we have an anti-fermion $i$ and fermion $j$ in the initial state and they approach along the $z$ axis, in the positive and negative directions respectively. If we choose $k_0$ to be a unit vector in the $z$ direction, i.e.
\be
k_0 = \big(1,0,0,1 \big),
\ee
then for momenta\footnote{$\beta = \big(1 - \frac{m^2}{E^2} \big)^{1/2}$}
\bea
\label{eq:pi} p_i &=& \big(E,0,0,\beta E \big), \\
\label{eq:pj} p_j &=& \big(E,0,0,-\beta E \big),
\eea
we have the following polarization vectors:
\bea
\sigma_i ^{\mu} &=& \frac{1}{m_i} \big(-E\beta,0,0,-E\big), \\
\sigma_j ^{\mu} &=& \frac{1}{m_j} \big(E\beta,0,0,-E\big).
\eea
If we recall that $m_i$ is negative because $i$ is an antiparticle, then we see that each polarization vector points in the same direction as the corresponding momentum, so that the spinors $u^{\lambda}(p)$ are indeed helicity eigenstates for this choice of $k_0$. However, choosing $k_0$ to be parallel to one of the particle's momenta results, in the massless limit, in the denominators of products such as that in \eqref{round bracket} vanishing. By being careful to take the limit algebraically this does not present a problem.\footnote{If we take $k_0^\mu = (1,0,\sin\theta,\cos\theta)$, then for the momenta defined in \eqref{eq:pi} and \eqref{eq:pj}, with $m_j = -m_i = m$, we have $(ij) = -2m\beta\cos\theta (1-\beta^2\cos^2\theta)^{-1/2}$. Thus $(ij) \sim O(m)$ as $m \to 0$ except if $\theta = 0^\circ$ when $(ij) \sim O(E)$.} But it should be noted that in such cases products like $(ij)$ do not necessarily vanish in the massless limit. We can sidestep this issue by choosing a different $k_0$, though we could not then talk of the helicity of the fermion.

\subsection{Example:  $\bar{q} q \to g g$}
To demonstrate the use of the massive spinor products described in the previous section we calculate the helicity amplitudes $M^{\lambda_1 \lambda_2 \lambda_3 \lambda_4}$ for the simple QCD process $\bar{q}^{\lambda_1}(p_1) \ q^{\lambda_2}(p_2) \to g^{\lambda_3}(p_3) \ g^{\lambda_4}(p_4)$. The $\lambda_1,\lambda_2 = \pm$ labels on the quarks refer to their spin polarizations in the sense already indicated. If we choose $k_0$ appropriately then they can be thought of as helicity labels. We will evaluate the partial (colour) amplitudes for the above scattering process, i.e. we consider contributions only from those diagrams with a particular ordering of the external gluons. The full colour-summed amplitudes can then be recovered by inserting appropriate colour factors, as described in Appendix B.

We first consider the $M^{+-+-}$ partial amplitude, for which there are two Feynman diagrams, shown in Figure~1. We will express them in terms of massive spinor products. For the slashed gluon polarization vectors we use
\bea
\s{\epsilon}^+(p,k) &=& \sqrt{2} \ \frac{u_+ (k) \bar{u}_+ (p) + u_- (p) \bar{u}_- (k)}{\la k p \ra}, \\
\s{\epsilon}^-(p,k) &=& \sqrt{2} \ \frac{u_+ (p) \bar{u}_+ (k) + u_- (k) \bar{u}_- (p)}{[pk]},
\eea
\begin{figure} \label{QCD_pmpm}
\centering
    \subfigure[Diagram A]{
    \label{fig:subfig:a}
    \includegraphics[width=2.3in]{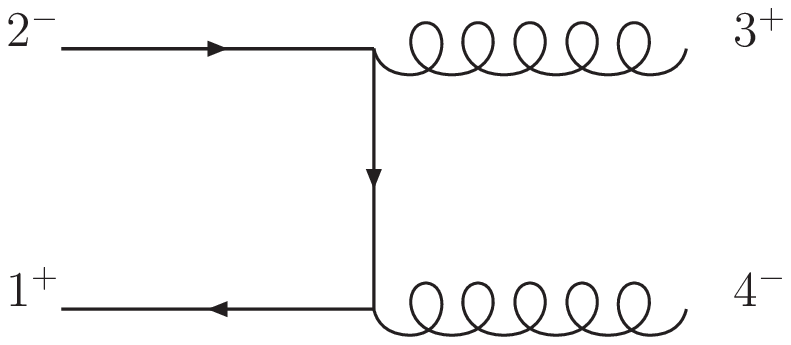}}
    \subfigure[Diagram B]{
    \label{fig:subfig:b}
    \includegraphics[width=2.3in]{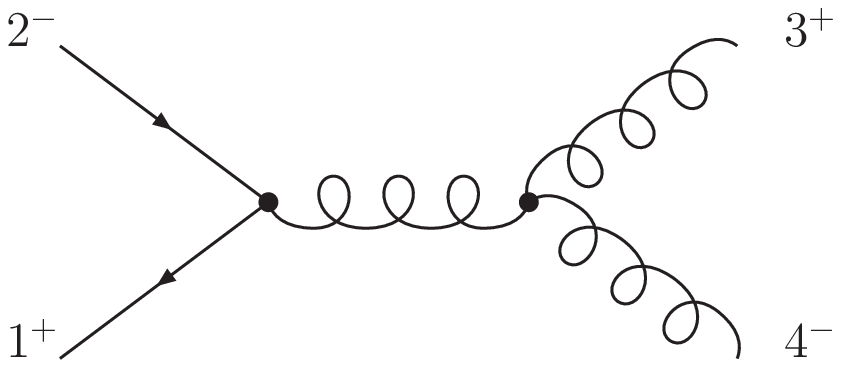}}
\caption{Diagrams contributing to the colour-ordered partial amplitude for the process $\bar{q}^+ (p_1) q^-(p_2) \to g^+(p_3) g^-(p_4)$.}
\end{figure}
where $k$ is a (null) reference vector which may be chosen separately for each gluon. Different choices of reference vector amount to working in different gauges. The choice $k_3 = p_4$ and $k_4 = p_3$ is particularly convenient in this context, as Diagram B vanishes in this gauge.
We have for the other diagram
\be \label{feynpmpm}
\bar{u}^+(p_1) \ \frac{\s{\epsilon}^-(p_4)}{\sqrt{2}} \ \frac{\s{p_2} - \s{p_3} + m}{(p_2 - p_3)^2 - m^2} \ \frac{\s{\epsilon}^+(p_3)}{\sqrt{2}} \ u^-(p_2),
\ee
which simplifies easily to
\be
\bar{u}^+(p_3) \s{p_2} u^+(p_4) \ \frac{\bar{u}^+(p_1) \big[u^-(p_3) \bar{u}^-(p_4) + u^+(p_4) \bar{u}^+(p_3)\big] u^-(p_2)}{4 \ p_3 \cdot p_4 \ p_4 \cdot p_1}, \\
\ee
so that
\be
M^{+-+-} =[3|2|4\ra \frac{[13](42) + (14)[32]}{4 \ p_3 \cdot p_4 \ p_4 \cdot p_1}.
\ee
As promised, we are left with an expression for the amplitude in terms of vector products and massive spinor products.

We next consider the other $M^{\lambda_1 \lambda_2+-}$ amplitudes. It is interesting to note that these are directly obtained from the $M^{+-+-}$ amplitude simply by changing the type of certain brackets. Thus
\bea
M^{+++-} &=& [3|2|4\ra \frac{[13]\la42\ra + (14)(32)}{4 \ p_3 \cdot p_4 \ p_4 \cdot p_1},\\
M^{--+-} &=& [3|2|4\ra \frac{(13)(42) + \la14\ra[32]}{4 \ p_3 \cdot p_4 \ p_4 \cdot p_1}, \\
M^{-++-} &=& [3|2|4\ra \frac{(13)\la42\ra + \la14\ra(32)}{4 \ p_3 \cdot p_4 \ p_4 \cdot p_1}.
\eea
Those amplitudes where the gluons have helicities $(- \ +)$ can be obtained directly from the ones above by complex conjugation.

 Let us now examine the case where the gluons have the same helicity. By direct calculation we find
\be
M^{--++} = m[43] \frac{\la13\ra(42) - \la14\ra(32)}{\la34\ra^2 \ 2 \ p_4 \cdot p_1}.
\ee
from which we deduce
\bea
M^{++++} &=& m[43] \frac{(13)\la42\ra - (14)\la32\ra}{\la34\ra^2 \ 2 \ p_4 \cdot p_1},\\
M^{+-++} &=& 0, \\
M^{-+++} &=& m[43] \frac{\la13\ra\la42\ra - \la14\ra\la32\ra}{\la34\ra^2 \ 2 \ p_4 \cdot p_1},\\
\label{glupp} &=& \frac{m[34]\la12\ra}{\la34\ra \ 2 \ p_4 \cdot p_1},
\eea
where in the last line we have used the Schouten identity. The amplitudes with two negative helicity gluons are obtained via complex conjugation. There are several interesting things to note about these results. First, the amplitude $M^{+-++}$ vanishes (for any choice of $k_0$) because of the identity\footnote{See Appendix A for a list of identities and notation.} $(13)(42) - (14)(32) = 0$. Second, when $k_0$ is parallel to the line of approach of the fermions (i.e. when we work in the helicity basis) then the product $\la12\ra$, and hence $M^{-+++}$, vanishes.

We have verified that when squared and summed over spins and colours, the set of $2 \to 2$ scattering amplitudes given above matches the well-known result (see for example Ref.~\cite{ESW}) calculated using Feynman diagrams and `trace technology', namely
\be
\sum_{\mbox{\tiny{colours}}} \sum_{\mbox{\tiny{spins}}} |M|^2 = 256 \ \left(\frac{1}{6\tau_1 \tau_2} -\frac{3}{8}\right) \left(\tau_1^2 + \tau_2^2 + \rho - \frac{\rho^2}{4\tau_1 \tau_2}\right),
\ee
where
\be
\tau_1 = \frac{2p_1 \cdot p_3}{s}, \hspace{5mm} \tau_2 = \frac{2p_1 \cdot p_4}{s}, \hspace{5mm} \rho=\frac{4m^2}{s}, \hspace{5mm} s=(p_1+p_2)^2.
\ee

Finally, the $m\to 0$ behaviour of the spin amplitudes can easily be read off from the expressions given above. For example, if $E$ denotes the typical scale of the $2\to 2$ scattering\footnote{We explicitly exclude zero angle scattering.}, then in the $m/E \to 0$ limit we have
\bea
M^{++ \pm \mp},\  M^{-- \pm \mp} &\sim & O(1), \nonumber \\
M^{+- \pm \mp},\  M^{-+ \mp \pm} &\sim & O(m/E), \nonumber\\
M^{++ \pm \pm},\  M^{-- \mp \mp} &\sim & O(m^2/E^2), \nonumber\\
M^{+---},\  M^{-+++} & \sim & O(m/E), \nonumber\\
M^{+-++},\  M^{-+--} & = & 0.
\label{eq:summary}
\eea
Note that in deriving these results we have assumed that $k_0$ is {\it not} directed along any of the particle momenta, so that all $(ij)$ spinor products are $O(m)$ in the $m\to 0$ limit, and $\la ij\ra$, $[ij]$ products are $O(E)$. If on the other hand we choose the (fermion) helicity basis by taking $k_0$ in the direction of (say) $p_1$, then
\eqref{eq:summary} becomes
\bea
M^{+- \pm \mp},\ M^{-+ \mp \pm} &\sim & O(1), \nonumber \\
M^{+- \pm \pm},\ M^{-+ \mp \mp} &  =  & 0, \nonumber\\
M^{++ \pm \mp},\ M^{-- \mp \pm} &\sim & O(m/E), \nonumber\\
M^{++++},\  M^{----} & \sim & O(m/E), \nonumber\\
M^{--++},\  M^{++--} & \sim & O(m^3/E^3) .
\label{eq:helsummary}
\eea

\section{BCFW Recursion}

In Ref.~\cite{BCF} Britto, Cachazo and Feng introduced new recursion relations for amplitudes involving gluons. The recursion involved \emph{on-shell} amplitudes with momenta shifted by a complex amount. Later \cite{BCFW}, the same authors with Witten gave an impressively simple and general proof of the recursion relations. They have since been successfully applied to amplitudes involving fermions \cite{LuoWen,KJO} and gravitons \cite{gravity}. Risager \cite{Risager} has demonstrated how they are related to the earlier `MHV rules', providing a proof of the latter simply using BCF recursion. There has also been much progress at 1-loop level \cite{loops1,loops2,loops3,loops4}, which has dovetailed nicely with the earlier unitarity work \cite{unitarity}.

We begin by choosing two (massless) particles $i$ and $j$ whose slashed\footnote{$\s{p} = \gamma^{\mu}p_{\mu}$} momenta we shift as follows,
\bea
\nonumber \s{p}_i \to \hat{\s{p}_i} &=& \s{p}_i + z\s{\eta}, \\
\label{shift} \s{p}_j \to \hat{\s{p}_j} &=& \s{p}_j - z\s{\eta},
\eea
where
\be
\s{\eta} = u^+(p_j) \bar{u}^+(p_i) + u^-(p_i) \bar{u}^-(p_j)
\ee
is such that both $p_i$ and $p_j$ remain on-shell. Using the familiar spin-sum condition,
\be
\sum_{\lambda} u^{\lambda}(p) \ \bar{u}^{\lambda}(p) = \s{p}
\ee
we can re-express the shift \eqref{shift} as a shift of spinors:
\bea \label{spinorshifts}
u^+(p_i) &\to& u^+(\widehat{p_i}) = u^+(p_i) + z \ u^+(p_j)  \\
\bar{u}^-(p_i) &\to& \bar{u}^-(\widehat{p_i})= \bar{u}^-(p_i)  + z \ \bar{u}^-(p_j) \\
\bar{u}^+(p_j) &\to& \bar{u}^+(\widehat{p_j})= \bar{u}^+(p_j) - z \ u^+(p_i) \\
\label{spinorshifts4} u^-(p_j) &\to& u^-(\widehat{p_j})= u^-(p_j) - z \ u^-(p_i).
\eea
In the Weyl spinor notation we are shifting $\lambda_i$ and $\tilde{\lambda}_j$. For massless particles, Dirac 4-spinors are effectively two copies of a Weyl 2-spinor, hence the four shifts of \eqref{spinorshifts}--\eqref{spinorshifts4}. Notice that there is no symmetry between $i$ and $j$ --- they are treated differently.

The amplitude is now a complex function of the parameter $z$. What the authors of \cite{BCFW} showed was that we can use the analytic properties of the amplitude as a function of $z$ to glean information about the physical case $z=0$. In particular, we get a recursion relation, which can be stated as
\be \label{recrelation}
A_n = \sum_{\mbox{\tiny{partitions}}} \sum_{\mbox{\tiny{s}}} A_{L}(\widehat{p}_i,\widehat{P}^{-s}) \ \frac{1}{P^2 - m_P^2} \ A_{R}(-\widehat{P}^{s},\widehat{p}_j).
\ee
where the hatted quantities are the shifted momenta. In fact, this is only valid if the helicities of the marked particles are chosen appropriately. The crucial property which must be retained if \eqref{recrelation} is to hold is that the shifted amplitude must vanish in the limit $z \to \infty$. There are rules \cite{BCFW,LuoWen,Badger1,Badger2} detailing which marking prescriptions are permitted in different cases. For our purposes, we will be on safe ground if the shifted gluons have helicites $(h^i, h^j) = (+,-)$ or $(\pm,\pm)$.

This method of calculation is particularly efficient because much of the computational complexity encountered in a Feynman diagram calculation is avoided since the lower point amplitudes $A_L$ and $A_R$ can be maximally simplified \emph{before} being inserted in \eqref{recrelation}.

The sum is over all partitions of the particles into a `left' group and a `right' group, subject to the requirement that particles $i$ and $j$ are on opposite sides of the divide. The sum over $s$ is a sum over the spins of the internal particle. Each diagram is associated with a particular value for the complex parameter $z$, which can be found via the condition that the internal momentum $\widehat{P}$ is on-shell. Note that $\widehat{P}$ is always a function of $z$ because of the restriction that the marked particles $i$ and $j$ appear on opposite sides of the divide.

One useful point to note in practice is that three-point gluon vertices vanish for certain marking choices. In particular, for the $j$ side of the diagram a gluon vertex with helicites $(++-)$ vanishes, as does the combination $(--+)$ on the $i$ side. This was pointed out in Ref.~\cite{BCF}.

We will be concerned in this work with the process $\bar{q} q \to ggg$, and so will encounter recursive diagrams connected by an internal fermion, the propagator of which is, in this formalism, the same as that of a scalar. Following Ref.~\cite{Badger2}, we `strip' fermions from the lower point amplitudes which feed the recursion and write
\bea
A_n &=& \sum_{\mbox{\tiny{partitions}}} \sum_{\mbox{\tiny{s}}} A_{L}(\widehat{p}_i,\widehat{P}^*) \ \frac{u_s(\widehat{P}) \bar{u}_s(\widehat{P})}{P^2 - m_P^2} \ A_{R}(-\widehat{P}^*,\widehat{p}_j), \\
&=& \sum_{\mbox{\tiny{partitions}}} A_{L}(\widehat{p}_i,\widehat{P}^*) \ \frac{\s{\widehat{P}} + m_P}{P^2 - m_P^2} \ A_{R}(-\widehat{P}^*,\widehat{p}_j).
\eea
where $P^*$ shows that the amplitude has been stripped of this external spinor wave-function. By way of example, let us reconsider the process $\bar{q}^+_1 q^-_2 \to g^+_3 g^-_4$. We mark the gluons such that $i=3$ and $j=4$. Then there is one recursive diagram,
\be \label{bcf+-+-}
\bar{u}^+(p_1) \frac{\s{\epsilon}^-(\widehat{p_4})}{\sqrt{2}} \ \frac{\s{p_2} - \s{\widehat{p_3}} + m}{(p_2 - p_3)^2 - m^2} \ \frac{\s{\epsilon}^+(\widehat{p_3})}{\sqrt{2}} u^-(p_2).
\ee
With the shifts we have chosen, the hats on the polarization vectors can be removed. The shifted part of the internal propagator is killed by either of the polarization vectors. So in fact all the hats can be removed in \eqref{bcf+-+-}, which is then identical to the Feynman diagram expression \eqref{feynpmpm}.

\section{$\bar{q} q \to 3g$ from BCFW Recursion}

The four-point amplitudes we derived in Section 2 are in such a form that it is trivial to strip a fermion off in the manner described above. This means that they are particularly convenient for use in BCFW recursion. Consider the process $\bar{q}^+_1 q^-_2 \to g^+_3 g^-_4 g^+_5$, for which there are three recursive diagrams, shown in Fig. 2. We choose the marking prescription $i=3$, $j=4$.

\begin{figure} \label{QCD_pmpmp}
\centering
    \subfigure[Diagram A]{
    \label{fig:subfig:c}
    \includegraphics[width=2.0in]{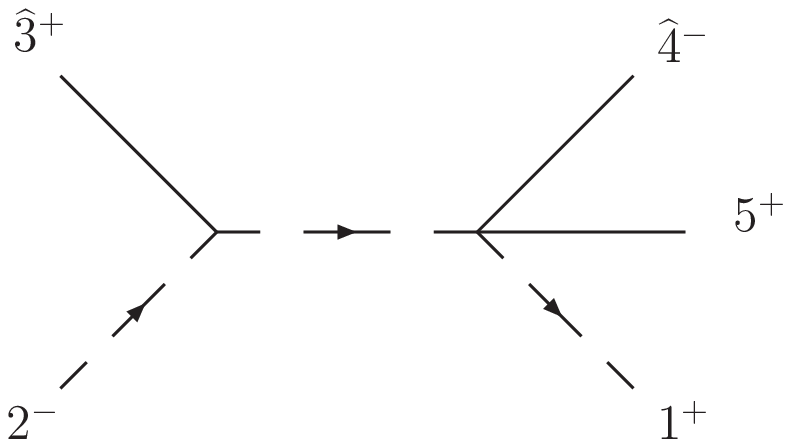}}
    \subfigure[Diagram B]{
    \label{fig:subfig:d}
    \includegraphics[width=2.0in]{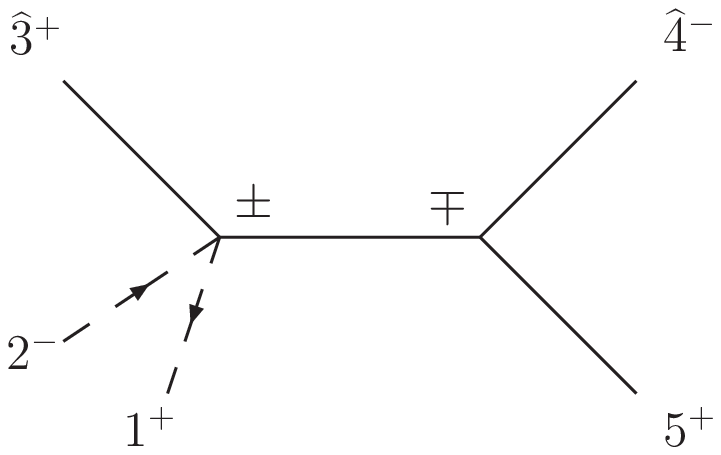}}
\caption{Recursive diagrams contributing to $\bar{q}^+ (p_1) q^-(p_2) \to g^+(p_3) g^-(p_4) g^+(p_5)$.}
\end{figure}

The two diagrams with internal gluons both vanish, due to the vanishing of $M^{+-++}$ and the vanishing of the $(++-)$ gluon vertex with the shifts we have chosen. For the remaining diagram we use
\be
M^{+--+} = -[4|1|3\ra \frac{(13)[42] + [14](32)}{4 \ p_3 \cdot p_4 \ p_4 \cdot p_1},
\ee
and strip the fermion $u^-(p_2)$, leaving
\be
M^{+\bullet-+} = -[4|1|3\ra \ \bar{u}^+(p_1) \ \frac{u^+(p_3) \bar{u}^+(p_4) + u^-(p_4) \bar{u}^-(p_3)}{4 \ p_3 \cdot p_4 \ p_4 \cdot p_1}.
\ee
After the appropriate relabelling this can be used in Diagram A, which can then be written
\bea
\nonumber A &=& -[5|1|\widehat{4}\ra \ \bar{u}^+(p_1) \ \frac{u^+(\widehat{p_4}) \bar{u}^+(p_5) + u^-(p_5) \bar{u}^-(\widehat{p_4}) }{4 \ p_5 \cdot \widehat{p_4} \ p_5 \cdot p_1} \ \frac{(\s{p}_2 - \s{\widehat{p}}_3 + m)}{(p_2-p_3)^2 - m^2} \\
&\times& \frac{\s{\epsilon}^+(\widehat{p_3})}{\sqrt{2}} u^-(p_2).
\eea
Due to our choice of marking, all the hats in the numerator can be removed. The shifted part of the propagator is killed by the gluon polarization vector. We are left with
\be \label{pmpmp}
M^{+-+-+} = [5|1|4\ra \left[ \frac{ m(14)(42)[53] + [15](42)[3|2|4\ra - (14)[32][5|1|4\ra}{8 \ p_5 \cdot p_1 \ p_2 \cdot p_3 \ \widehat{p}_4 \cdot p_5 \ \la43\ra} \right].
\ee
We can work out $z$ from the requirement that $\widehat{P}^2 = (p_2 - \widehat{p}_3)^2 = m^2$, and find
\be
z = \frac{-2 \ p_2 \cdot p_3}{[3|2|4\ra}.
\ee
The product $\widehat{p}_4 \cdot p_5$ is then
\bea
\widehat{p}_4 \cdot p_5&=&(p_4 - z \ \eta) \cdot p_5 \\
&=& p_4 \cdot p_5 + \frac{p_2 \cdot p_3}{[3|2|4\ra} [3|5|4\ra.
\eea
The result \eqref{pmpmp} is much more compact than the expression obtained from a Feynman diagram calculation, with which it agrees. See Section 5 for the Feynman results for this process in terms of massive spinor products.

\subsection{Results for Helicity Conserving Amplitudes}

Here we give all the helicity conserving QCD amplitudes for $\bar{q} q \to ggg$. By helicity conserving we mean those amplitudes where the spin polarizations of the fermions are $+-$, in the sense described in Section 2. Whether these labels actually correspond physically to helicity depends on the choice of $k_0$. We choose to mark adjacent gluons, so that each amplitude has contributing recursive diagrams of the form of Fig. 2a and Fig. 2b, that is, we have a diagram with an internal fermion and a diagram with an internal gluon. The vanishing of the $2 \to 2$ amplitude $M^{+-++}$ simplifies those cases where there is a majority of positive helicity gluons. In particular, the diagrams with an internal gluon vanish. In the remaining cases, we evaluate such diagrams in the same way as in Ref.~\cite{BCF}, using identities such as
\bea
[A \widehat{P}] &=& \frac{[A | P | i\ra}{\la \widehat{P} i \ra}, \\
\la \widehat{P} B \ra &=& \frac{[j | P | B \ra}{[j \widehat{P}]},
\eea
with $i$ and $j$ as in \eqref{shift}. These identities hold only when the marked particles $i$ and $j$ are massless.

The results presented here are valid for arbitrary spin polarizations. Choosing a polarization basis amounts to choosing the vector $k_0$, and when this is done the expressions below will simplify. In the helicity basis for example, in which we choose $k_0$ to be parallel to the line of approach of the fermions, the building block $M^{+---}$ vanishes. This causes the first term in each of the mostly-minus amplitudes below to vanish.

\be
M^{+-+-+} = [5|1|4\ra \left[ \frac{ m(14)(42)[53] + [15](42)[3|2|4\ra - (14)[32][5|1|4\ra}{8 \ p_5 \cdot p_1 \ p_2 \cdot p_3 \ \widehat{p}_4 \cdot p_5 \ \la43\ra} \right]
\ee
where $i=3$, $j=4$ and
\be
\nonumber \widehat{p}_4 \cdot p_5 = p_4 \cdot p_5 + \frac{p_2 \cdot p_3}{[3|2|4\ra} [3|5|4\ra.
\ee
\vspace{8mm}
\be
\begin{split}
&M^{+-++-} = [\widehat{4}|1|5\ra \times \\
  &\left[ \frac{ m[1\widehat{4}][32]\la54\ra + [1\widehat{4}](42)[3|2|5\ra + 2 p_5 \cdot p_1 (15)[32] + m(15)(42)[43]}{8 \ p_2 \cdot p_3 \ \widehat{p}_4 \cdot p_5  \ p_5 \cdot p_1 \la43\ra} \right]
\end{split}
\ee
where $i=3$, $j=4$ and
\be
\nonumber \widehat{p_4} \cdot p_5 = p_4 \cdot p_5 + \frac{ p_2 \cdot p_3 }{[3|2|4\ra}[3|5|4\ra, \hspace{11mm}
 |\widehat{4}] = |4] - \frac{(-2 \ p_2 \cdot p_3)}{[3|2|4\ra} |3].
\ee
\vspace{8mm}

\be
\begin{split}
&M^{+--++} = [\widehat{4}|2|3\ra \times \\
  &\left[ \frac{ m[\widehat{4}2][15]\la43\ra - [\widehat{4}2](14)[5|1|3\ra - 2 p_2 \cdot p_3 [15](32) + m(14)(32)[54]}{8 \ p_2 \cdot p_3 \ p_3 \cdot \widehat{p}_4 \ p_5 \cdot p_1 \la54\ra} \right]
\end{split}
\ee
where $i=5$, $j=4$ and
\be
\nonumber  p_3 \cdot \widehat{p}_4 = p_3 \cdot p_4 + \frac{ p_1 \cdot p_5 }{[5|1|4\ra}[5|3|4\ra, \hspace{11mm}
 |\widehat{4}] = |4] - \frac{(-2 \  p_1 \cdot p_5)}{[5|1|4\ra} |5].
\ee

\vspace{8mm}

\be
\begin{split}
&M^{+-+--} = \frac{m [21]\la45\ra^3}{\la34\ra\big[\la35\ra 2 p_5 \cdot p_1 + \la34\ra[4|2|5\ra\big](p_1 + p_2)^2} \ +  \\
 [3|2|\widehat{4}\ra
& \left[ \frac{ m(\widehat{4}2)(15)[43] - (\widehat{4}2)[14][3|1|5\ra - 2 p_2 \cdot p_3 (15)[32] + m[14][32]\la54\ra}{8 \ p_2 \cdot p_3 \ p_3 \cdot \widehat{p}_4 \ p_5 \cdot p_1 [54]} \right]
\end{split}
\ee
where $i=4$, $j=5$, and
\be
\nonumber  p_3 \cdot \widehat{p}_4 = p_3 \cdot p_4 + \frac{ p_1 \cdot p_5 }{[4|1|5\ra}[4|3|5\ra, \hspace{11mm}
 (\widehat{4} 2) = (42) + \frac{(2  p_1 \cdot p_5)}{[4|1|5\ra} (52).
\ee

\vspace{8mm}

\bea
M^{+--+-} &=& \frac{m [21]\la53\ra^4}{\sqrt{2} \ \la43\ra\la45\ra\big[\la53\ra 2 \ p_2 \cdot p_3 + \la54\ra[4|2|3\ra\big](p_1 + p_2)^2} \\
\nonumber &+& [4|1|5\ra \left[ \frac{ m[14][42]\la53\ra + (15)[42][4|2|3\ra - [14](32)[4|1|5\ra}{8 \ p_5 \cdot p_1 \ p_2 \cdot p_3 \  \widehat{p}_4  \cdot p_5 [34]} \right]
\eea
where $i=4$, $j=3$, and
\be
\nonumber  \widehat{p}_4 \cdot p_5 = p_4 \cdot p_5 + \frac{ p_2 \cdot p_3 }{[4|2|3\ra}[4|5|3\ra.
\ee

\vspace{8mm}
\bea
&&M^{+---+} = \frac{m [21]\la43\ra^3}{\sqrt{2} \ \la45\ra\big[\la53\ra 2 \ p_2 \cdot p_3 + \la54\ra[4|2|3\ra\big](p_1 + p_2)^2} \ + \\
\nonumber && [5|1|\widehat{4}\ra \times  \left[ \frac{ m(1\widehat{4})(32)[54] + (1\widehat{4})[42][5|2|3\ra + 2 p_5 \cdot p_1 [15](32) + m[15][42]\la43\ra}{8 \ p_2 \cdot p_3 \ \widehat{p}_4 \cdot p_5  \ p_5 \cdot p_1 [34]} \right]
\eea
where $i=4$, $j=3$, and
\bea
\nonumber  \widehat{p}_4 \cdot p_5 &=& p_4 \cdot p_5 + \frac{ p_2 \cdot p_3 }{[4|2|3\ra}[4|5|3\ra, \hspace{11mm}
 (1\widehat{4} ) = (14) + \frac{(2 \  p_2 \cdot p_3)}{[4|2|3\ra} (13), \\
\nonumber |\widehat{4}\ra & =& |4\ra + \frac{2 \ p_2 \cdot p_3}{[4|2|3\ra} |3\ra
\eea
\vspace{8mm}

\be
M^{+----} = \frac{m \la5 \widehat{4} \ra [4|2|3\ra [12][45]}{4 \ [45]^2 \  p_5 \cdot p_1 \ p_2 \cdot p_3 \ [34]}
\ee
where
\be
\nonumber \la 5 \widehat{4} \ra= \la 54 \ra + \frac{2p_2 \cdot p_3}{[4|2|3\ra}\la53\ra.
\ee

\be
M^{+-+++} = 0
\ee
The amplitudes with fermion helicities $-+$ can be obtained from those above by complex conjugation.
\subsection{Results for Helicity Flip Amplitudes}

We now consider the helicity flip amplitudes. These have fermion spin polarization labels $\pm \pm$. In this case we are unable to evaluate diagrams with an internal gluon due to products such as $(\widehat{P} k)$ where $k$ is massive. For those amplitudes in which all gluons have the same helicity, the internal gluon diagrams vanish anyway:

\be
M^{++---} = m\la\widehat{4}3\ra \left[ \frac{ m(15)(42)[43] + [14](32)[4|1|5\ra - [14](42)[3|1|5\ra}{-4 \ p_5 \cdot p_1 \ p_2 \cdot p_3 [34]^2 [54]} \right]
\ee
where $i=4$, $j=5$ and
\be
\nonumber |\widehat{4}\ra = |4\ra + \frac{2 \ p_5 \cdot p_1}{[4|1|5\ra} |5\ra.
\ee

\vspace{8mm}
\be
M^{+++++} = m[5\widehat{4}] \left[ \frac{ m(14)(32)\la54\ra + (14)\la42\ra [3|2|5\ra - (15)\la42\ra[3|2|4\ra}{-4 \ p_5 \cdot p_1 \ p_2 \cdot p_3 \la45\ra^2 \la43\ra} \right]
\ee
where $i=3$, $j=4$ and
\be
\nonumber |\widehat{4}] = |4] + \frac{2 \ p_2 \cdot p_3}{[3|2|4\ra} |3].
\ee
\\
The amplitudes $M^{--+++}$ and $M^{-----}$ are obtained from those above by complex conjugation. For the remaining amplitudes we resort to Feynman diagrams.

\section{Feynman Results}

Here we give results for $\bar{q}q \to ggg$ derived from Feynman rules. Note that in a given amplitude all the helicities can be flipped by complex conjugation.

\bea
\nonumber M^{+--+-} &=& -[4|2|3\ra \left[ \frac{ m[14][42]\la53\ra + (15)[42][4|2|3\ra - [14](32)[4|1|5\ra}{8 \ p_5 \cdot p_1 \ p_2 \cdot p_3 \ p_3 \cdot p_4 \ [54]} \right] \\
\nonumber \\
\nonumber &+&\la35\ra \bigg[\frac{m[14][42]\la53\ra + (15)[42][4|2|3\ra - [14](32)[4|1|5\ra }{8 \ p_2 \cdot p_3 \ p_3 \cdot p_4 \ p_4 \cdot p_5}\bigg] \\
\nonumber \\
 &+&\frac{\la35\ra^2}{\la34\ra\la45\ra(p_1 + p_2)^2}\bigg[\frac{[14](32) + (13)[42]}{[54]} + \frac{[14](52) + (15)[42]}{[34]}\bigg].
\eea

\bea
\nonumber M^{+-++-} &=& -[4|1|5\ra \left[ \frac{-m(15)(52)[43] + (15)[32][4|1|5\ra - [14](52)[3|2|5\ra}{8 \ p_5 \cdot p_1 \ p_2 \cdot p_3 \ p_4 \cdot p_5 \ \la53\ra} \right] \\
\nonumber \\
\nonumber  &+& [43][4|1|5\ra\bigg[\frac{[14](52) + (15)[42]}{4 \ p_5 \cdot p_1 \ p_3 \cdot p_4 \la 53\ra [54] }\bigg] \\
\nonumber \\
\nonumber &-& [43]\bigg[\frac{[14](52)[3|1|5\ra + (15)[32][4|1|5\ra - m(15)(52)[43]}{4 \ p_5 \cdot p_1 \ p_3 \cdot p_4 \ p_4 \cdot p_5}\bigg] \\
\nonumber \\
&-&\frac{[43]^2\la35\ra}{2 \ p_3 \cdot p_4 [54](p_1 + p_2)^2}\bigg[\frac{[14](52) + (15)[42]}{\la53\ra} + \frac{[13](52) + (15)[32]}{\la54\ra]}\bigg].
\eea

\bea
\nonumber M^{+-+--} &=& [3|2|4\ra \left[ \frac{m[13][32]\la54\ra - [13](42)[3|1|5\ra + (15)[32][3|2|4\ra}{-8 \ p_5 \cdot p_1 \ p_2 \cdot p_3 \ p_3 \cdot p_4 \ [53]} \right] \\
\nonumber \\
\nonumber &+& \la45\ra[3|2|4\ra\bigg[\frac{[13](42) + (14)[32]}{4 \ p_2 \cdot p_3 \ p_4 \cdot p_5 \la 43\ra [35] }\bigg] \\
\nonumber \\
\nonumber &+& \la45\ra \bigg[\frac{[13](42)[3|2|5\ra + (15)[32][3|2|4\ra + m[13][32]\la54\ra]}{8 \ p_2 \cdot p_3 \ p_3 \cdot p_4 \ p_4 \cdot p_5}\bigg] \\
\nonumber \\
&+&\frac{\la45\ra^2[53]}{2 \ p_4 \cdot p_5 \la43\ra(p_1 + p_2)^2}\bigg[\frac{[13](52) + (15)[32]}{[43]} + \frac{[13](42) + (14)[32]}{[53]}\bigg].
\eea

The corresponding helicity flip amplitudes can be obtained from these simply by altering the types of brackets. For example, suppose we wish to extract $M^{---+-}$ from $M^{+--+-}$ given above. We can achieve this by changing brackets as follows:
\bea
[1 k] &\to& (1k), \\
(1k) &\to& \la 1 k \ra,
\eea
where $k$ is massless. Sandwich products such as $[4|1|5\ra$ are not changed.
This transformation results in
\bea
\nonumber M^{---+-} &=& -[4|2|3\ra \left[ \frac{ m(14)[42]\la53\ra + \la15\ra[42][4|2|3\ra - (14)(32)[4|1|5\ra}{8 \ p_5 \cdot p_1 \ p_2 \cdot p_3 \ p_3 \cdot p_4 \ [54]} \right] \\
\nonumber \\
\nonumber &+&\la35\ra \bigg[\frac{m(14)[42]\la53\ra + \la15\ra[42][4|2|3\ra - (14)(32)[4|1|5\ra }{8 \ p_2 \cdot p_3 \ p_3 \cdot p_4 \ p_4 \cdot p_5}\bigg] \\
\nonumber \\
 &+&\frac{\la35\ra^2}{\la34\ra\la45\ra(p_1 + p_2)^2}\bigg[\frac{(14)(32) + \la13\ra[42]}{[54]} + \frac{(14)(52) + \la15\ra[42]}{[34]}\bigg].
\eea
Other amplitudes can be found by analogous bracket alterations.

\section{Summary}

We have calculated all the partial spin amplitudes for the $\bar{q}q \to ggg$ scattering process where $q$ is a massive fermion.
For most of the partial amplitudes we were able to use the BCFW recursion relations to obtain fairly compact expressions. This was achieved by following the idea of Ref.~\cite{Badger2} of stripping lower point amplitudes of their external fermion wavefunctions before inserting them into the recursion. We used a particular representation of massive spinors, along the lines of the appendix of Ref.~\cite{KS2}, to define massive spinor products. In this method information regarding the polarization of the fermion spins is contained in the definition of the spinor products, rather than explicitly in the amplitude.

We derived very compact results for the {\em helicity conserving} partial amplitudes. This simplicity can be attributed to the vanishing of certain $2 \to 2$ scattering amplitudes, which meant that in some cases we were able to avoid diagrams with internal gluons. We were unable to treat the {\em helicity flip} amplitudes in the same way (except for the case where all the gluon helicities are the same), since we were unable to evaluate the corresponding recursive diagrams with internal gluons. For these amplitudes we instead provided expressions derived from Feynman diagrams, also in terms of massive spinor products.

These results represent an interesting test of the BCFW recursion relations \cite{BCF,BCFW}. The massive spinor products we used are well suited to calculations with massive fermions. Application of these techniques to higher order processes with massive fermions, such as $\bar q q \to gggg$, should be possible though would be accompanied by an increase in complexity.

\vskip 1cm
\noindent{\bf Acknowledgements} \\\\
KJO acknowledges the award of a PPARC studentship. We are grateful to Valery Khoze for useful discussions.

\newpage
\section*{Appendices}
\appendix

\section{Notation and Conventions}

We have used products of Dirac spinors
\be
[ij]=\bar{u}^+(i) u^-(j), \hskip 1cm  \la ij\ra=\bar{u}^-(i) u^+(j), \hskip 1cm (ij)=\bar{u}^{\pm}(i) u^{\pm}(j)
\ee
with massive $p_i$,$p_j$. To evaluate these we use two arbitrary four vectors $k_0$ and $k_1$, such that
\be
k_0^2 = 0, \hskip 1cm k_1^2=-1, \hskip 1cm k_0 \cdot k_1 = 0.
\ee
Then
\bea
[ij] &=& \frac{(p_i \cdot k_0) (p_j \cdot k_1) - (p_j \cdot k_0)(p_i \cdot k_1) - i \epsilon_{\mu \nu \rho \sigma}k_0^\mu p_i^\nu p_j^\rho k_1^\sigma}{\sqrt{(p_i\cdot k_0)(p_j \cdot k_0)}} \\
\la ij \ra &=& \frac{(p_j \cdot k_0)(p_i \cdot k_1) -(p_j \cdot k_1) (p_i \cdot k_0) - i \epsilon_{\mu \nu \rho \sigma}k_0^\mu p_i^\nu p_j^\rho k_1^\sigma}{\sqrt{(p_i\cdot k_0)(p_j \cdot k_0)}}\\
(ij) &=& m_i \left(\frac{p_j \cdot k_0}{p_i \cdot k_0}\right)^{\frac{1}{2}} + i \leftrightarrow j
\eea
where $m_i$ is negative if $i$ is an antiparticle. Different choices of $k_0$ correspond to different choices of the quantization axis of a massive fermion's spin, as described in Section 2.

We use the notation
\bea
\bar{u}^+(i) \ \s{p} \ u^+(j) &=& [i|p|j\ra = [i p]\la pj \ra + (ip)(pj), \\
\bar{u}^-(i) \ \s{p} \ u^-(j) &=& \la i|p|j] = \la i p \ra [pj] + (ip)(pj).
\eea
Whereas for massless vectors $k_i$,$k_j$ we have the familiar relation $2k_i \cdot k_j = \la ij \ra [ji]$, in the massive case this is extended to
\be
2 p_i \cdot p_j = \la ij \ra [ji] + (ij)^2.
\ee
For any massive $i$,$j$ and massless $k,l$ we have
\bea
(ik)(jl) &=& (il)(jk), \\
(ik)[li] + [ik](li) &=& m_i [lk], \\
\bar{u}^{\pm}(p_k) \ \s{p_i} \ u^{\mp}(p_l) &=& 0
\eea
The Schouten identity holds,
\be
\langle a \ b \rangle \langle c \ d \rangle + \langle a \ c \rangle \langle d \ b \rangle + \langle a \ d \rangle \langle b \ c \rangle=0.
\ee
For gluon polarization vectors we use
\bea
\epsilon^{+}_{\mu}(p,k) &=&  \frac{\bar{u}^{-}(k) \ \gamma_{\mu} \ u^{-}(p)}{\sqrt{2} \ \la kp \ra }, \\
\epsilon^{-}_{\mu}(p,k) &=&  \frac{\bar{u}^{+}(k) \ \gamma_{\mu} \ u^{+}(p)}{\sqrt{2} \ [ pk ] },
\eea
which take the slashed form
\bea
\s{\epsilon}^+(p,k) &=& \sqrt{2} \ \frac{u_+ (k) \bar{u}_+ (p) + u_- (p) \bar{u}_- (k)}{\la k p \ra}, \\
\s{\epsilon}^-(p,k) &=& \sqrt{2} \ \frac{u_+ (p) \bar{u}_+ (k) + u_- (k) \bar{u}_- (p)}{[pk]}.
\eea
We use a shorthand form for the amplitude in which we display only the helicities of the particles involved. So for example,
\be
M(\bar{q}^+_1 , q^-_2 \ ; \ 3^+, 4^-, 5^+)  \sim M^{(+-+-+)}.
\ee

\section{Colour Decomposition}
The calculation of multi-parton scattering amplitudes in perturbative QCD becomes problematic very quickly as the number of partons increases, due to the sheer number of diagrams and the complicated gluon self-interactions. One technique to circumvent this is to split the set of all Feynman diagrams contributing to a particular amplitude into gauge invariant subsets. Then different gauges can be used in the evaluation of each subset. This simplifies the overall calculation considerably. Each subset of Feynman diagrams is called a \emph{partial amplitude}.
We use the normalization
\be
\mbox{Tr}(T^A T^B) = \delta_{ab}.
\ee
where the $T^i$ are matrices of the fundamental representation of $SU(3)$. This convention leads to colour-ordered Feynman rules as given in \cite{Dixon}. The colour decomposition for processes with a pair of quarks is then
\be
A(\bar{q},q \ ; \ g_1,g_2 \cdots g_n) = \sum_{\sigma} \big(T^{a_{\sigma(1)}} \dots T^{a_{\sigma(n)}}\big)_{ij} M(\bar{q},q \ ; \ g_{\sigma(1)},g_{\sigma(2)} \cdots g_{\sigma(n)}).
\ee
Here $\sigma$ is the set of all distinct cyclic orderings of the gluons. The colour information in a given amplitude is contained purely in the group theoretical prefactors, while all the kinematical information is contained in the partial amplitudes $M(\sigma)$. It is useful to note that amplitudes in QED can be obtained from amplitudes in QCD by replacing all the colour matrices $T^A$ with the identity matrix. For references and a more detailed description of colour decomposition, the reader is directed to Ref. \cite{Dixon}.

\end{document}